\documentclass{article}
\usepackage{spconf,amsmath,graphicx,bbm,array,dcolumn,epsfig,amsmath,amsfonts,yhmath,bm,amssymb,psfrag,color,soul,enumerate,stackengine,cite}
\usepackage[dvipsnames]{xcolor}
\usepackage{tcolorbox}
\usepackage[font=footnotesize,skip=7pt]{caption}
\usepackage[font=small,skip=0pt]{subcaption}
\usepackage[ruled,vlined]{algorithm2e}
\usepackage{comment}
\usepackage[lining]{ebgaramond}

\usepackage{tikz}

\definecolor{lgreen} {RGB}{180,210,100}
\definecolor{ngreen} {RGB}{98,158,31}
\definecolor{dgreen} {RGB}{78,138,21}
\definecolor{MLOWLSgreen} {RGB}{0,140,130}
\definecolor{SDPpurple} {RGB}{191,0,191}
\definecolor{lred}   {RGB}{220,0,0}
\definecolor{nred}   {RGB}{224,0,0}
\definecolor{bred}   {RGB}{200,20,20}
\definecolor{nblue}  {RGB}{28,130,185}
\definecolor{jblue}  {RGB}{20,50,100}

\newcommand*{\myfontb}{\fontfamily{lmr}\selectfont}

\newcommand {\myvec}[1] {{\mbox{\boldmath $#1$}}}
\newcommand {\mymat}[1]  {{\mbox{\boldmath $#1$}}}

\DeclareMathAlphabet      {\mathbfit}{OML}{cmm}{b}{it}

\AtBeginEnvironment{bmatrix}{\setlength{\arraycolsep}{4pt}}

\newcommand {\mSigma} {\mymat{\Sigma}}

\newcommand {\A} {\mymat{A}}

\newcommand {\hA} {\widehat{\A}}

\newcommand {\mGamma} {\mymat{\Gamma}}

\newcommand {\uhalf} {\myvec{\frac{1}{2}}}

\renewcommand {\H} {\mymat{H}}

\newcommand {\ubv} {\mybar{\myvec{v}}}
\newcommand {\V} {\mymat{V}}

\newcommand {\I} {\mymat{I}}

\newcommand {\ue} {\myvec{e}}
\newcommand {\ua} {\myvec{a}}

\newcommand {\ug} {\myvec{g}}

\newcommand {\uxi} {\myvec{\xi}}

\newcommand {\uv} {\myvec{v}}

\newcommand {\uo} {\myvec{0}}
\newcommand {\us} {\myvec{s}}

\newcommand {\ux} {\myvec{x}}

\newcommand {\uy} {\myvec{y}}
\newcommand {\uw} {\myvec{w}}

\newcommand {\uz} {\myvec{z}}

\newcommand {\Rset} {\mathbb{R}}

\newcommand {\Zset} {\mathbb{Z}}
\newcommand {\Eset} {\mathbb{E}}
\newcommand {\Nset} {\mathbb{N}}

\newcommand {\tps} {\rm{T}}

\DeclareMathOperator*{\argmin}{argmin}

\makeatletter
\newsavebox\myboxA
\newsavebox\myboxB
\newlength\mylenA

\newcommand*\mybar[2][0.75]{%
	\sbox{\myboxA}{$\m@th#2$}%
	\setbox\myboxB\null
	\ht\myboxB=\ht\myboxA%
	\dp\myboxB=\dp\myboxA%
	\wd\myboxB=#1\wd\myboxA
	\sbox\myboxB{$\m@th\overline{\copy\myboxB}$}
	\setlength\mylenA{\the\wd\myboxA}
	\addtolength\mylenA{-\the\wd\myboxB}%
	\ifdim\wd\myboxB<\wd\myboxA%
	\rlap{\hskip 0.5\mylenA\usebox\myboxB}{\usebox\myboxA}%
	\else
	\hskip -0.5\mylenA\rlap{\usebox\myboxA}{\hskip 0.5\mylenA\usebox\myboxB}%
	\fi}
\makeatother

\title{Blind Modulo Analog-to-Digital Conversion of Vector Processes}

\name{Amir Weiss$^{\star}$, Everest Huang$^\dagger$, Or Ordentlich$^{\diamondsuit}$ and Gregory W. Wornell$^{\star}$}

\address{
\begin{tabular}{ccc}
$^{\star}$Dept. of EECS &  $^\dagger$MIT Lincoln Laboratory & $^{\diamondsuit}$School of CSE \\
Massachusetts Institute of Technology & everest@ll.mit.edu  &Hebrew University of Jerusalem\\
\{amirwei,gww\}@mit.edu & \; & or.ordentlich@mail.huji.ac.il
\end{tabular}
\thanks{\scriptsize
This material is based upon work supported, in part, by the United States Air Force under Air Force Contract No.~FA8702-15-D-0001. Any opinions, findings, conclusions or recommendations expressed in this material are those of the author(s) and do not necessarily reflect the views of the United States Air Force. This work was also supported, in part, by ISF under Grant 1791/17, and NSF under Grant CCF-1717610.}
\thanks{\scriptsize DISTRIBUTION STATEMENT A. Approved for public release. Distribution is unlimited. \textcopyright\, 2021 Massachusetts Institute of Technology.
}
\thanks{\scriptsize Delivered to the U.S. Government with Unlimited Rights, as defined in DFARS Part 252.227-7013 or 7014 (Feb 2014). Notwithstanding any copyright notice, U.S.\ Government rights in this work are defined by DFARS 252.227-7013 or DFARS 252.227-7014 as detailed above. Use of this work other than as specifically authorized by the U.S.\ Government may violate any copyrights that exist in this work.}
}

\begin{document}
\ninept
\maketitle
\setlength{\abovedisplayskip}{5pt}
\setlength{\belowdisplayskip}{5pt}

\begin{abstract}
\small{In a growing number of applications, there is a need to digitize a (possibly high) number of correlated signals whose spectral characteristics are challenging for traditional analog-to-digital converters (ADCs). Examples, among others, include multiple-input multiple-output systems where the ADCs must acquire at once several signals at a very wide but sparsely and dynamically occupied bandwidth supporting diverse services. In such scenarios, the resolution requirements can be prohibitively high. As an alternative, the recently proposed \emph{modulo-ADC} architecture can in principle require dramatically fewer bits in the conversion to obtain the target fidelity, but requires that spatiotemporal information be known and explicitly taken into account by the analog and digital processing in the converter, which is frequently impractical. Building on our recent work, we address this limitation and develop a \emph{blind} version of the architecture that requires no such knowledge in the converter. In particular, it features an automatic modulo-level adjustment and a fully adaptive modulo-decoding mechanism, allowing it to asymptotically match the characteristics of the unknown input signal. Simulation results demonstrate the successful operation of the proposed algorithm.}
\end{abstract}

\vspace{-0.3cm}
\section{Introduction}\label{sec:intro}
\vspace{-0.2cm}
In a host of applications in communication and signal processing there is often a need to digitize highly correlated analog signals, where each signal, which in general may be temporally correlated in itself, is fed into a separate analog-to-digital converter (ADC). One representative example is in the context of a massive multiple-input multiple-output channels \cite{larsson2014massive,lu2014overview,marzetta2015massive}, where the number of antennas can be of the order of tens and even hundreds, whereas the number of users (/independent sources) it serves is moderate, making the signals received by the antennas highly correlated.

In such scenarios, when the signals are (naively) acquired by standard ADCs, there is a high degree of redundancy in the data. Naturally, this redundancy can be exploited in many ways, depending on the processing phase and the desired objective \cite{donoho2006compressed,zhang2011sparse,ahmed2019compressive,ahmed2019sub,shlezinger2019joint}. Particularly attractive is to utilize this redundancy as early as the acquisition phase, namely in the analog to digital conversion. By doing so, in principle, the signals could be digitized at the same fidelity using fewer bits, thus reducing power consumption, which in general grows exponentially with the number of bits \cite{walden1999analog}.

One possible approach to put this notion into practice is to use the recently proposed \emph{modulo ADCs}~\cite{ordentlich2018modulo,ordonez2021full}. A modulo ADC first folds each sample of the input process modulo $\Delta$, where $\Delta$ is a design parameter, and only then quantizes the result using a traditional uniform quantizer. The modulo operation limits the dynamic range of the signal to be quantized, which in turns results in a quantization error whose magnitude is proportional to $\Delta$, rather than to the dynamic range of the original, unfolded signal. In~\cite{ordentlich2018modulo} it is shown that the observed signal can be reliably unfolded, when the second-order statistics (SOSs) of the input signals are known, and $\Delta$ is set proportionally to the prediction error standard deviation. More recently, a blind mechanism for a single modulo ADC was proposed~\cite{weiss2021blind}, which adapts the effective modulo size (analogously to an AGC mechanism in a standard ADC \cite{sun2010automatic}) and learns the required SOSs of the input, while unwrapping the folded signal with the same reliability.

In this work, we extend \cite{weiss2021blind}, and develop a \emph{blind} mechanism for \emph{multiple} modulo ADCs working in parallel. Using spatiotemporal correlations of the observed signals, the proposed solution learns the inputs' underlying SOSs, adapts the effective modulo size, and for a given number of bits, considerably decreases the mean square error (MSE) distortion in the reconstruction of the input signals relative to standard ADCs.

\begin{figure}[t]
	\includegraphics[width=0.45\textwidth]{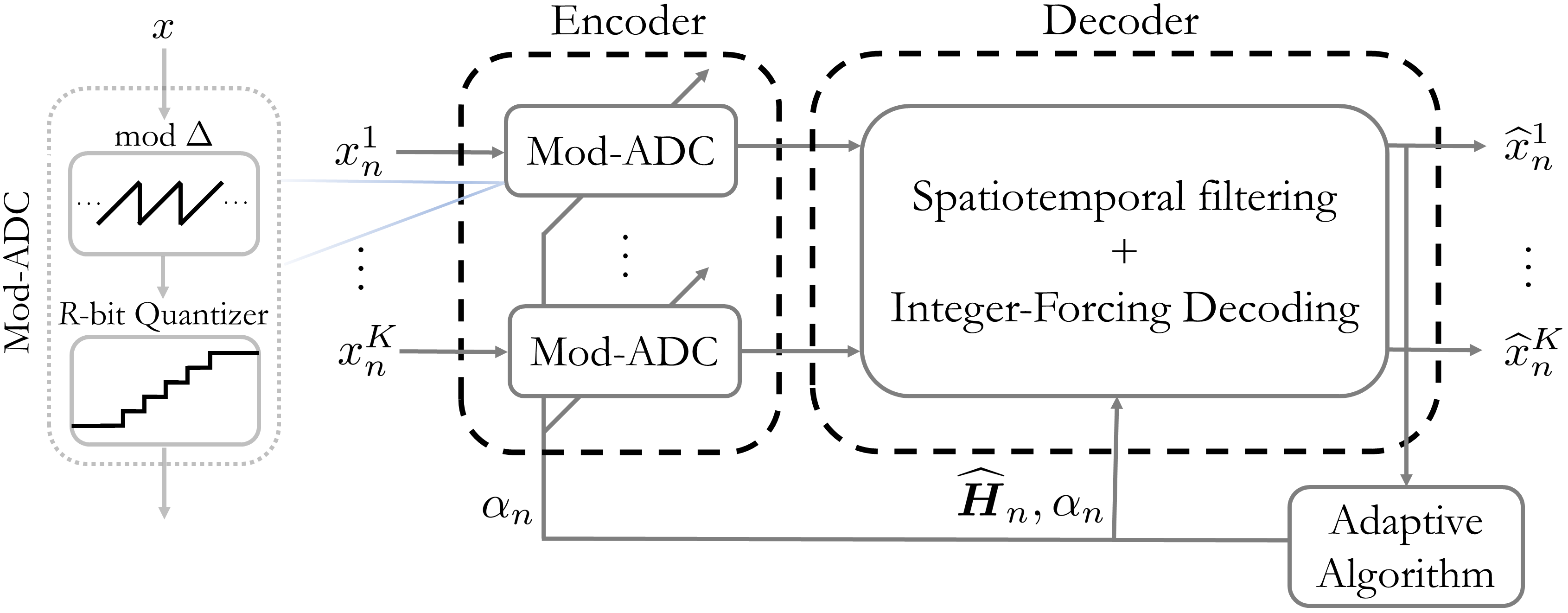}
	\centering\vspace{-0.1cm}
	\caption{A schematic high-level illustration of the proposed blind mod-ADCs.}
	\label{fig:blockblindModADC}\vspace{-0.7cm}
\end{figure}

\vspace{-0.35cm}
\section{Brief Review on Modulo ADCs}\label{sec:idealmoduloADC}
\vspace{-0.15cm}
For a positive number $\Delta\in\Rset^+$, we define
\begin{equation*}
[x]\;{\rm{mod}}\;\Delta\triangleq x-\Delta\cdot\left\lfloor \frac{x}{\Delta}\right\rfloor\in[0,\Delta), \quad \forall x\in\Rset,
\end{equation*}
as the $[\cdot]\;{\rm{mod}}\;\Delta$ operator, where $\left\lfloor x\right\rfloor$ is the floor operation, which returns the largest integer smaller than or equal to $x$. An $R$-bit modulo ADC with resolution parameter $\alpha$, termed $(R,\alpha)$ mod-ADC, computes
\begin{equation}\label{modADCdefinition}
[x]_{R,\alpha}\triangleq\left[\left\lfloor \alpha x\right\rfloor\right]\;{\rm{mod}}\;2^R\in\{0,1,\ldots,2^R-1\},
\end{equation}
and produces the binary representation of \eqref{modADCdefinition} as its output (Fig.\ \ref{fig:blockblindModADC}, left).

Using subtractive dithers \cite{lipshitz1992quantization}, an $(R,\alpha)$ mod-ADC can be modeled as a stochastic channel, whose output $y$ for an input $x$ is given by
\begin{equation}\label{modulorandomchannel}
y=\left[\alpha x+z\right]\;{\rm{mod}}\;2^R,
\end{equation}
where $z\sim{\rm{Unif}}\left((-1,0]\right)$. Since the modulo operation is a form of \emph{lossy} compression, it is generally impossible to recover the unfolded signal $(\alpha x+z)$ from its folded version $y$ \eqref{modulorandomchannel}. Nevertheless, under relatively mild conditions, when the input (possibly high-dimensional) signal is sufficiently temporally- and/or spatially-predictable, e.g., a correlated random vector-process \cite{ordentlich2018modulo} or a deterministic bandlimited signal \cite{bhandari2017unlimited,romanov2019above}, it is in fact possible to \emph{perfectly} recover the unfolded signal\footnote{With high probability for random signals, and to an arbitrary precision for deterministic bandlimited signals.} from its past samples and its current folded sample via causal processing \cite{ordentlich2016integer}.

More generally, consider $K$ parallel $(R,\alpha)$ mod-ADCs whose input signals $\{x_n^k\}$, collected into a vector $\ux_n\in\Rset^{K\times 1}$, are zero-mean jointly stationary processes, with known autocorrelation functions $\{R^{ij}_{x}[\ell]\triangleq\Eset\left[x^i_nx^j_{n-\ell}\right]\in\Rset\}$. The output of the $K$ mod-ADCs is then given by
\begin{equation*}
\uy_n=\left[\alpha \ux_n+\uz_n\right]\;{\rm{mod}}\;2^R\in\Rset^{K\times 1},\;\;\forall n\in\Nset^+,
\end{equation*}
where $\{z^k_n\sim {\rm{Unif}}((-1,0])\}$, modeling the quantization noises, are independent, identically distributed (i.i.d.) stochastic processes, and the modulo is elementwise. Further, define the unfolded quantized signal,
\begin{equation*}
\uv_n\triangleq\alpha \ux_n + \uz_n\in\Rset^{K\times 1},\;\;\forall n\in\Nset^+,
\end{equation*}
and assume that the decoder has access to $\{\uv_{n-1},\ldots,\uv_{n-p}\}$. Notice that once $\uv_n$ is recovered, $\ux_n$ is readily estimated as $\widehat{\ux}_n=(\uv_n+\uhalf)/\alpha$. Thus, we focus on recovering $\uv_n$ based on $\uy_n$ and $\uv_{[n]}\triangleq{\rm{vec}}\left(\V_n\right)\in\Rset^{Kp\times 1}$, where $\V_n\triangleq[\uv_{n-1} \cdots \uv_{n-p}]\in\Rset^{K\times p}$.
\begin{algorithm}[t]
	\nl Compute the Linear Minimum MSE (LMMSE) estimate of $\uv_n$ based on $\uv_{[n]}$ (denoting $\uhalf$ as the all-$\frac{1}{2}$ vector)
	{\setlength{\abovedisplayskip}{5pt}\setlength{\belowdisplayskip}{5pt}
	\begin{equation}\label{LMMSElengthp}
	\widehat{\uv}_{\text{\tiny LMMSE},n}^p=\H^p_{\text{opt}}\left(\uv_{[n]}+\uhalf\right)-\uhalf,
	\end{equation}}
	where $\H^p_{\text{opt}}\in\Rset^{K\times Kp}$ is the matrix filter corresponding the LMMSE predictor, computed based on $\{R^{ij}_x[\ell]\}$ and $\alpha$;\\
	\nl Denote $\A = [\ua_1 \cdots \ua_K]^{\tps}$, and solve [\emph{IF decoding matrix}]
	{\setlength{\abovedisplayskip}{2pt}\setlength{\belowdisplayskip}{2pt}
	\begin{equation}\label{oracelIFproblem}
	    \A \triangleq \argmin_{\substack{\bar{{\mathbf{A}}}\in\Zset^{K\times K} \\ \det(\bar{\mathbf{A}})\neq0}} \max_{k\in\{1,\ldots,K\}} \frac{1}{2}\log_2\left(\mybar{\ua}_k^{\tps}\mSigma_p\mybar{\ua}_k\right),
	\end{equation}}
	where $\mSigma_p$ \eqref{errorcovariancedef} is computed based on $\{R_x^{ij}[\ell]\}$ and $\alpha$;\\
	\nl Compute $\uw_{\text{\tiny LMMSE},n}=[\uy_n-\widehat{\uv}_{\text{\tiny LMMSE},n}^p]\;{\rm{mod}}\;2^R$;\\
	\nl Compute $\widetilde{\ug}_n\in\Rset^{K\times 1}$ and the estimated prediction error,
	{\setlength{\abovedisplayskip}{5pt}\setlength{\belowdisplayskip}{5pt}
	\begin{equation}\label{IFdecodingoracle}
	\hspace*{-0.2cm}\begin{cases}
	    g_n^k\triangleq \left[\ua_k^{\tps}\uw_{\text{\tiny LMMSE},n}\right]\,{\rm{mod}}\;2^R\\
	    \widetilde{g}_n^k\triangleq \left[g_n^k+\frac{1}{2}2^R\right]\,{\rm{mod}}\;2^R\hspace{-0.05cm}-\hspace{-0.05cm}\frac{1}{2}2^R
	\end{cases}\hspace{-0.4cm}\Rightarrow \widehat{\ue}_{\text{\tiny LMMSE},n}^{\,p}\hspace{-0.05cm}\triangleq\hspace{-0.05cm}\A^{-1}\widetilde{\ug}_n;
	\end{equation}}\\
	\nl Return $\widehat{\uv}_{\text{oracle},n}=\widehat{\uv}_{\text{\tiny LMMSE},n}^p+\widehat{\ue}_{\text{\tiny LMMSE},n}^{\,p}$.
	\caption{{\bf Oracle Spatiotemporal Modulo Unfolding} \newline $\widehat{\uv}_{\text{oracle},n}=\text{ \texttt{OracleSTModUnfold} }(\uy_n, \uv_{[n]}, \{R^{ij}_x[\ell]\}, \alpha, R)$\label{Algorithm1}}
\end{algorithm}
\setlength{\textfloatsep}{1pt}

The algorithm proposed in \cite{ordentlich2018modulo} for recovering $\uv_n$ with high probability (w.h.p.) when $\{\uv_{n-\ell}\}_{\ell=1}^p$ and $\{R^{ij}_x[\ell]\}$ are \emph{known}, here referred to as \emph{oracle} spatiotemporal modulo unfolding, is given in Algorithm \ref{Algorithm1}. The key idea behind this method is that the modulo operation essentially becomes invertible (w.h.p.) when statistical properties of the original (unfolded) signal and correlated measurements to the modulo-folded signal are available, provided that $\Delta$ is proportional to the standard deviation of the innovation process. In Algorithm \ref{Algorithm1}, unfolding of the coordinates of $\uv_n$ is done by using spatiotemporal (causal) linear filtering, followed by \emph{integer-forcing} (IF) decoding \cite{sakzad2013integer,ding2015exact,ordentlich2016integer,huang2017transceiver}, which exploits spatial correlations in the prediction error vector. In this regard, solving \eqref{oracelIFproblem} generally requires a complexity exponential in $K$. However, the optimal integer matrix $\A$ needs to be computed only once for the pair $\{\mSigma_p, \alpha\}$, where
\begin{equation}\label{errorcovariancedef}
\mSigma_p\triangleq\Eset\left[\ue_{\text{\tiny LMMSE},n}^{\,p}\left(\ue_{\text{\tiny LMMSE},n}^{\,p}\right)^{\tps}\right]\in\Rset^{K\times K}
\end{equation}
is the covariance matrix of the error $\ue_{\text{\tiny LMMSE},n}^{\,p}\triangleq\uv_n-\widehat{\uv}_{\text{\tiny LMMSE},n}^p$ in optimal linear prediction, and $\widehat{\uv}_{\text{\tiny LMMSE},n}^p$ is defined in \eqref{LMMSElengthp}. Moreover, the LLL algorithm \cite{lenstra1982factoring} provides a sub-optimal $\A$ with a complexity $\mathcal{O}(\mathrm{poly}(K))$.

Of course, in practice, devices such as ADCs usually operate under dynamic conditions, giving rise to a wide range of possible inputs with unknown characteristics, and must still maintain proper operation. Therefore, a significant step towards implementing mod-ADCs for real-life applications can be made by relaxing the (too restrictive) assumption that $\{R^{ij}_x[\ell]\}$ are all known. Building on our recent work, in which a single blind mod-ADC was developed based on temporal correlations, the goal of this work is to take that important step for the general case of $K$ parallel mod-ADCs, exploiting both temporal and spatial correlations.
\begin{algorithm}[t]
	\nl Compute the linear estimate of $\uv_n$ based on $\ubv_{[n]}$
	{\setlength{\abovedisplayskip}{5pt}\setlength{\belowdisplayskip}{5pt}
	\begin{equation}\label{linearestimation}
	\widehat{\uv}_{n}^p=\widehat{\H}^p_{n}\widehat{\ubv}_{[n]}-\uhalf;
	\end{equation}}\\
	\nl Denote $\hA = [\widehat{\ua}_1 \cdots \widehat{\ua}_K]^{\tps}$, and solve
	{\setlength{\abovedisplayskip}{2pt}\setlength{\belowdisplayskip}{2pt}
	\begin{equation}\label{blindIFdecod}
	    \hA \triangleq \argmin_{\substack{\bar{{\mathbf{A}}}\in\Zset^{K\times K} \\ \det(\bar{\mathbf{A}})\neq0}} \max_{k\in\{1,\ldots,K\}} \frac{1}{2}\log_2\left(\mybar{\ua}_k^{\tps}\widehat{\mSigma}_p\mybar{\ua}_k\right);
	\end{equation}}\\
	\nl Compute $\uw_n = [\uy_n-\widehat{\uv}_n^p]\;{\rm{mod}}\;2^R$;\\
	\nl Compute $\widehat{\widetilde{\ug}}_n\in\Rset^{K\times 1}$ and the estimated prediction error,
	{\setlength{\abovedisplayskip}{5pt}\setlength{\belowdisplayskip}{5pt}
	\begin{equation}\label{IFdecodingoracleblind}
	\hspace*{-0.5cm}\begin{cases}
	    \widehat{g}_n^k\triangleq \left[\widehat{\ua}_k^{\tps}\uw_{n}\right]\,{\rm{mod}}\;2^R\\
	    \widehat{\widetilde{g}}_n^k\triangleq \left[\widehat{g}_n^k+\frac{1}{2}2^R\right]\,{\rm{mod}}\;2^R\hspace{-0.05cm}-\hspace{-0.05cm}\frac{1}{2}2^R
	\end{cases}\hspace{-0.4cm}\Rightarrow \widehat{\ue}_{n}^{\,p}\triangleq\hA^{-1}\widehat{\widetilde{\ug}}_n;
	\end{equation}}\\
	\nl Return $\widehat{\uv}_{n}=\widehat{\uv}_{n}^p+\widehat{\ue}_{n}^{\,p}$, $\widehat{\ue}_{n}^{\,p}$.
	\caption{{\bf Blind Spatiotemporal Modulo Unfolding} \newline $\widehat{\uv}_n,\,\widehat{\ue}_{n}^{\,p}\hspace{-0.05cm}=\hspace{-0.05cm}{\text{ \texttt{BlindSTModUnfold} }}(y_n, \widehat{\ubv}_{[n]}, \widehat{\H}^p_n, \widehat{\mSigma}_p, \alpha_n, R)$\label{Algorithm2}}
\end{algorithm}
\setlength{\textfloatsep}{1pt}

\vspace{-0.35cm}
\section{Problem Formulation}\label{sec:problemformulation}
\vspace{-0.2cm}
Consider $K$ parallel $(R,\alpha_n)$ mod-ADC as described in the previous section, with a fixed modulo range $\Delta=2^R$, but an \emph{adaptable}, possibly \emph{time-varying} resolution parameter $\alpha_n\in\Rset^+$. The mod-ADCs are fed with the input signal $\{\ux_n\triangleq \ux(nT_s)\}_{n\in \Nset^+}$, acquired by sampling the analog, continuous-time signal $\ux(t)$ every $T_s=f_s^{-1}$ seconds. We assume that $\{x^k_n\}$ are zero-mean jointly stationary stochastic processes with \emph{unknown} correlation functions $\{R^{ij}_{x}[\ell]\}$. The observed, distorted signals at the output of the $K$ mod-ADCs are given (in vector form) by
\begin{equation*}
\uy_n=[\alpha_n \ux_n+\uz_n]\;{\rm{mod}}\;2^R, \; \forall n\in\Nset^+,
\end{equation*}
where, as before, the quantization noise processes $\{z^k_n\hspace{-0.05cm}\sim\hspace{-0.05cm}{\rm{Unif}}((-1,0])\}$ are i.i.d. Further, we redefine the unfolded quantized signal,
\begin{equation}\label{defofv}
\uv_n\triangleq\alpha_n \ux_n + \uz_n, \; \forall n\in\Nset^+,
\end{equation}
which, in general, is no longer stationary. Nonetheless, whenever $\alpha_n$ is held fixed, $\uv_n$ can be regarded as stationary on the respective time period.

As explained above, the goal in this context is to estimate the input $\ux_n$ as accurately as possible based on the observed sequence $\{\uy_n\}$ at the output of the mod-ADCs using a causal system. However, since $\uv_n$ is merely a scaled version of $\ux_n$ contaminated by white noise \eqref{defofv}, the problem essentially boils down to recovering $\uv_n$, and is stated as follows.

\noindent\textbf{Problem Statement:} {\myfontb For a fixed number of bits $R$, design an adaptive mechanism for estimating $\{\ux_n\}$ from the mod-ADCs' outputs with the lowest possible MSE distortion, \emph{without prior knowledge on the correlation functions $\{R^{ij}_x[\ell]\}$}.
}

\noindent The above is interpreted as designing an update mechanism for causally maximizing the resolution parameter $\alpha_n$, while still allowing for reliable recovery of $\uv_n$ from $\left\{\uy_k\right\}_{k\leq n}$, and also design the recovery mechanism.

As explained in Section \ref{sec:idealmoduloADC}, provided $\uv_n$ is exactly recovered w.h.p., i.e., $\widehat{\uv}_n\overset{\text{w.h.p.}}{=}\uv_n$, the input signal is readily estimated as $\widehat{\ux}_n\triangleq\frac{\widehat{\uv}_n+\uhalf}{\alpha_n}$, 
where $\alpha_n$ is a known parameter, and $\uhalf$ is to compensate for the quantization noise (non-zero) expectation $\Eset[\uz_n]=-\uhalf$ (also in \eqref{LMMSElengthp} and \eqref{linearestimation}).

\vspace{-0.35cm}
\section{Blind Modulo ADC of Vector Processes}\label{sec:proposedsolution}
\vspace{-0.25cm}
We now present the blind mod-ADC algorithm, which simultaneously estimates the input $\ux_n$, while learning the (possibly time-varying) SOSs of the unfolded signal \eqref{defofv}, necessary for estimation of $\ux_n$. The structure of the proposed architecture is depicted in Fig.\ \ref{fig:blockblindModADC}. In contrast to the oracle mod-ADC (\textit{cf}.\ Fig.\ 6 in \cite{ordentlich2018modulo}), here the encoder and decoder are adaptive, and vary with time according to the statistical properties of the input.

The underlying concept of our approach is the following. For a fixed resolution parameter $\alpha_n$, given that at any time instance $n$ the unfolded signal $\uv_n$ can be \emph{exactly} recovered, we may estimate the optimal matrix filter $\H^p_{\text{opt}}$, corresponding to the LMMSE predictor of $\uv_n$ \eqref{LMMSElengthp} based on the previous samples $\V_n$. This can be done using the least mean squares (LMS) algorithm \cite{haykin2003least}, which converges\footnote{In the mean sense, under mild conditions stated explicitly in the sequel.} to $\H^p_{\text{opt}}$. Upon convergence, $\alpha_n$ can be slightly increased, and as long as the spatiotemporal prediction is sufficiently accurate to allow successful IF decoding as in \eqref{IFdecodingoracle}, $\uv_n$ could still be recovered using the same technique as in Algorithm \ref{Algorithm1}. Fixing $\alpha_n$ again to its new value, the new optimal filter is learned using LMS. The process is repeated until the resolution parameter $\alpha_n$ is sufficiently large.

Conceptually, only appropriate initial conditions and sufficiently smooth transitions from one resolution level to another are required for convergence of this process. Once these are fulfilled, we attain successful steady state operation of the $K$ blind mod-ADCs. We now formalize this concept, and develop the desired adaptive mechanism.

\vspace{-0.35cm}
\subsection{Phase 1: Initialization and Learning the Optimal Decoding}\label{subsec:phase1initialization}
\vspace{-0.1cm}
We begin with a ``small" initial value for the resolution parameter, $\alpha_0$, that ensures a degenerated modulo operation, i.e., $\widetilde{\uy}_n=\uv_n$, where
\begin{equation*}
\widetilde{\uy}_n\triangleq\left(\left[\uy_n+\tfrac{1}{2}2^R\right]\;{\rm{mod}}\;2^R\right)-\tfrac{1}{2}2^R,
\end{equation*}
such that $\widetilde{y}_n$ is the ``modulo-shifted" version of $y_n$. For purposes that will become clear in the sequel, we further define for convenience
\begin{equation}\label{normalizedv}
\mybar{\uv}_n\triangleq\left(\uv_n+\uhalf\right)/\alpha_n= \ux_n + \left(\uz_n+\uhalf\right)/\alpha_n,
\end{equation}
the ``$\alpha_n$-standardized" version of $\uv_n$. Note that $\Eset\left[\mybar{\uv}_n\right]=\uo$, and the covariance of $\mybar{\uv}_n$ is dominated by the covariance of $\ux_n$ when $\alpha_n$ is large.

Assuming $\widetilde{y}_n=v_n$ as long as $\alpha_n=\alpha_0$ is fixed, the optimal filter for prediction of $\uv_n$ \eqref{defofv} based on $\ubv_{[n]}\triangleq{\rm{vec}}\left([\mybar{\uv}_{n-1}\cdots\mybar{\uv}_{n-p}]\right)\in\Rset^{Kp\times1}$, can be estimated with the LMS algorithm \cite{haykin2003least}, via the updates
\begin{equation*}
\widehat{\H}^p_n=\widehat{\H}^p_{n-1}+\mu\cdot\ue_n^p\ubv_{[n]}^{\tps}\in\Rset^{K\times Kp}.
\end{equation*}
Here, $\widehat{\H}^p_n$ is the filter used in Algorithm \ref{Algorithm2} for the prediction of $\uv_n$, $\mu$ is the learning rate (or step size), and 
\begin{equation}\label{linearpredictionerror}
\ue_n^p\triangleq \uv_n-\widehat{\uv}_n^p\in\Rset^{K\times 1}
\end{equation}
is the prediction error of the linear predictor $\widehat{\uv}_n^p$ as in \eqref{linearestimation}. In addition, rather than using $\{\uv_n\}$ \eqref{defofv}, we use the standardized process $\{\mybar{\uv}_n\}$ as the observations in \eqref{linearestimation}, since as the adaptive process evolves and $\alpha_n$ increases, the SOSs of $\mybar{\uv}_n$ become less affected by $\alpha_n$, alleviating the LMS. Details regarding the convergence of the LMS algorithm, as well as the proper setting of the step size $\mu$ which guarantees this convergence, are omitted due to space limitation (see \cite{weiss2021blind} for the same guiding principles). From now on, assume that $\mu$ is chosen so as to ensure that \cite{feuer1985convergence},
\begin{equation}\label{meanconvergencetoopt}
\alpha_n=\alpha_0:\;\lim_{n\to\infty} \Eset\left[\widehat{\H}^p_n\right]=\H^p_{\text{opt}}.
\end{equation}

After enough iterations, since we assume that $\alpha_0$ is sufficiently small to ensure that $\widehat{\uv}_n=\uv_n$ for every $n$ during initialization, which gives us access to $\widehat{\ue}_n^p=\ue_n^p$ \eqref{linearpredictionerror}, the filter $\widehat{\H}^p_n$ will approximately converge to an unbiased estimate of $\H^p_{\text{opt}}$, as in \eqref{meanconvergencetoopt}. Further, during the initialization we can compute the empirical covariance $\widehat{\mSigma}_p\triangleq\frac{1}{n}\sum_{\ell=1}^n\widehat{\ue}_{\ell}^p\left(\widehat{\ue}_{\ell}^p\right)^{\tps}$, and use it for IF decoding as in \eqref{blindIFdecod}. Assuming $\mu$ is sufficiently small, the MSE of $\widehat{\uv}^p_n$ will approximately converge to the MSE of $\widehat{\uv}_{\text{\tiny LMMSE},n}^p$ (with $\alpha_n$ replacing $\alpha$, according to the definition \eqref{defofv}),
\begin{equation}\label{defMSEoflinearest}
\exists N_0:\forall n>N_0:\Eset\left[\ue_n^p\left(\ue_n^p\right)^{\tps}\right]\underset{\mu\ll1}{\approx}\mSigma_{p}.
\end{equation}
Once \eqref{defMSEoflinearest} is accurate enough and $\widehat{\mSigma}_p\approx\mSigma_p$, by \cite[Prop. 5]{ordentlich2018modulo}, we have
\begin{equation}\label{defOLofaaptive}
\begin{aligned}
\Pr(\mathcal{E}_{\tiny {\rm{OL}}_n})&\triangleq\Pr\left(\widehat{\uv}_n\neq\uv_n\right)\approx\left.\Pr\left(\widehat{\uv}_{\text{\tiny oracle},n}\neq\uv_n\right)\right|_{\text{\boldmath$A$}=\widehat{\text{\boldmath$A$}}}\\
&\left.\leq2K\exp\left\{-\frac{3}{2}2^{2(R-R_{\text{IF}}(\mathbf{A}))}\right\}\right|_{\text{\boldmath$A$}=\widehat{\text{\boldmath$A$}}},
\end{aligned}
\end{equation}
where $\mathcal{E}_{\tiny {\rm{OL}}_n}$ is referred to as the overload event,
\begin{equation*}
R_{\text{IF}}(\A)\triangleq \max_{k\in\{1,\ldots,K\}} \frac{1}{2}\log_2\left(\mybar{\ua}_k^{\tps}\mSigma_p\mybar{\ua}_k\right)\triangleq\frac{1}{2}\log_2\left(\sigma_{\max}^2\right),
\end{equation*}
and $\uv_n$ is recovered w.h.p. The approximation in \eqref{defOLofaaptive} is mostly due to estimation errors; a sharper analysis is omitted due to lack of space. Having learned the optimal decoding, we can increase the resolution parameter $\alpha_n$, and use the quantizers' output bits to a finer description of inputs.

\vspace{-0.35cm}
\subsection{Phase 2: Updating the Resolution Parameter}\label{phase2adaptation}
\vspace{-0.15cm}
As shown in \cite[Prop. 5]{ordentlich2018modulo}, the no overload event $\mathcal{E}^*_{\tiny \mybar{{\rm{OL}}}_n}\triangleq\{\widehat{\uv}_{\text{\tiny oracle},n}^p=\uv_n\}$ of the oracle mod-ADCs is equivalent to $\{\max_k|g_n^k|<\frac{\Delta}{2}\}$, where $\ug_n^k\triangleq\A\ue_n^p$. Therefore, in order to increase $\alpha_n$, we must somehow detect that the filter $\H^p_n$ already approximates the optimal one well enough, such that the \emph{transformed prediction errors} $\{g_n^k\}$ are sufficiently small with respect to the dynamic range $\Delta$. In this case, a small change in the resolution would not affect our ability to recover $\uv_n$ w.h.p. Due to space limitation, we refer the reader to \cite{weiss2021blind}, Subsection IV-B, for a detailed justification for a single mod-ADC, based on similar principles.

Since the resolution update rule is a user-controlled system parameter, it can be easily set to be small enough, so as to maintain sufficiently small transformed prediction errors, and safely continue recovering $\uv_n$ w.h.p.\ after the update. It remains only to make sure that before the resolution parameter is increased, we arrive at an intermediate steady state, wherein $\mathcal{E}_{\tiny \mybar{{\rm{OL}}}_n}$ occurs w.h.p. A plausible way of achieving this, is to update $\alpha_n$ only when the maximum variance of the transformed prediction errors is much smaller than $\Delta$, i.e., when $\kappa\cdot\sigma_{\max}<\frac{\Delta}{2}$ for some large enough non-negative constant $\kappa\in\Rset^+$. If the prediction errors were Gaussian, this would provide a direct proxy for the overload probability, and increasing $\kappa$ would exponentially decrease the overload probability. This provides the conditions to re-learn the optimal filter corresponding to the LMMSE predictor of $\uv_n$ with the updated resolution $\alpha_{n}$.

Recall that in practice, though, since $\{R^{ij}_x[\ell]\}$ are unknown, $\sigma_{\max}$ is clearly not known as well. Nevertheless, since $\widehat{\uv}_n\overset{\text{w.h.p.}}{=}\uv_n\Leftrightarrow\widehat{\ug}_n\overset{\text{w.h.p.}}{=}\ug_n$ throughout the adaptive process, we can estimate $\sigma_{\max}$ online by
{\setlength{\abovedisplayskip}{3pt}\setlength{\belowdisplayskip}{3pt}
\begin{equation*}
{\widehat{\sigma}}^2_{k,n}\triangleq\frac{1}{L_s}\sum_{\ell=0}^{L_s-1}\left({\widehat{g}}^k_{n-\ell}\right)^2,\; {\widehat{\sigma}}_{\max,n}\triangleq\max_{k\in\{1,\ldots,K\}}\sqrt{{\widehat{\sigma}}^2_{k,n}}.
\end{equation*}}
where $L_s\in\Nset^+$ is a moving-average window length, and is also set to be the minimal time stabilization interval wherein $\alpha_n$ must be kept fixed after its last update. Thus, we increase $\alpha_{n}$ only when $\mathbbm{1}_{\uparrow\alpha,n}=1$, where
\begin{equation}\label{indicatorkappastd}
\mathbbm{1}_{\uparrow\alpha,n} \triangleq \begin{cases}
1, & \kappa\cdot\widehat{\sigma}_{\max,n}<\frac{\Delta}{2}\\
0, & \kappa\cdot\widehat{\sigma}_{\max,n}>\frac{\Delta}{2}
\end{cases}.
\end{equation}
Whenever $\mathbbm{1}_{\uparrow\alpha,n}=0$, we infer that the prediction errors are not satisfactorily small. In these cases, we decrease the resolution so as to resort to a state where $\uv_n$ is again recovered w.h.p, which allows the LMS to converge to the desired filter, and then safely increase the resolution again.

Upon updating $\alpha_n$, we also appropriately update the filter $\widehat{\H}^p_n$, since the input signal $\ubv_n$ is scaled with $\alpha_n$ as well \eqref{normalizedv}. Thus, we update
\begin{align*}
\alpha_{n+1}&=\left[\mathbbm{1}_{\uparrow\alpha,n}\cdot\delta^{-1}_{\alpha}+\left(\mathbbm{1}_{\uparrow\alpha,n}-1\right)\cdot\delta_{\alpha}\right]\alpha_{n},\\
\widehat{\H}^p_{n+1}&=\left[\mathbbm{1}_{\uparrow\alpha,n}\cdot\delta^{-1}_{\alpha}+\left(\mathbbm{1}_{\uparrow\alpha,n}-1\right)\cdot\delta_{\alpha}\right]\widehat{\H}^p_{n},
\end{align*}
where $\delta_{\alpha}\in(0,1)$ is the update resolution parameter.

It is now straightforward to generalize this adaptive process, since, conceptually, we now only need to repeatedly execute the properly interlaced Phase 1 and Phase 2. In the repeated Phase 1, the ``initial" values for the filter and resolution parameter would be the corresponding values of the previous time step. Additionally, $\ubv_{[n]}$ will be replaced by $\widehat{\mybar{\uv}}_{[n]}\triangleq{\mathrm{vec}}\left(\left[\widehat{\mybar{\uv}}_{n-1}\, \cdots\, \widehat{\mybar{\uv}}_{n-p}\right]\right)\in\Rset^{Kp\times1}$, whose vector-elements\footnote{We use in \eqref{defofvbarest} the (seemingly redundant) notation $\widehat{\mybar{\uv}}_{n}$ rather than $\widehat{\ux}_n$, since in this context we trying to perfectly recover $\mybar{\uv}_n$, rather than to estimate $x_n$.}
\begin{equation}\label{defofvbarest}
\widehat{\mybar{\uv}}_{n}\triangleq\widehat{\ux}_n=\left(\widehat{\uv}_n+\uhalf\right)/\alpha_{n}
\end{equation}
are $\{\widehat{\mybar{\uv}}_{i}\overset{\text{w.h.p.}}{=}\mybar{\uv}_{i}\}_{i=n-1}^{n-p}$. The repeated Phase 2 would then be executed after (at least) $L_s$ time steps with the updated resolution.

\vspace{-0.35cm}
\subsection{Detecting an Overload, and Preventing Error Propagation}\label{subsec:errorpropagation}
\vspace{-0.1cm}
Given there is no overload, which holds w.h.p.\ when $\kappa$ is large enough, alternating between these two phases leads to convergence near the limiting fixed point $\kappa\cdot\sigma_{\max,n}=\frac{\Delta}{2}$, as in \eqref{indicatorkappastd}, up to small fluctuations due to the limited-resolution adaptation step $\delta_{\alpha}$ and estimation errors in $\widehat{\sigma}_{\max,n}$. However, the overload event still has non-zero probability, and must be taken into account. Indeed, whenever $\mathcal{E}_{\tiny {\rm{OL}}_n}=\{\widehat{\uv}_n\neq \uv_n\}$ occurs, an extremely fast error propagation process begins. To detect such errors and prevent the consequent error propagation, we propose a generalized version of the detector developed in \cite{weiss2021blind} (Subsection IV-C therein),
\begin{equation}\label{ourdetector}
\widehat{\mathbbm{1}}_{\mathcal{E}_{\tiny {\rm{OL}}_n}}\hspace{-0.1cm}\triangleq \begin{cases}
1, & \bigcup\limits_{k=1}^K\left\{|\widehat{\mybar{v}}^k_n|>\hspace{-0.05cm}\sqrt{2\widehat{\sigma}_{\bar{v}^k,n}^2\log(n)}\right\}\\
0, & \text{otherwise}
\end{cases}, \, \forall n\geq N_s,
\end{equation}
such that \eqref{ourdetector} detects an outlier with a large magnitude w.h.p.\footnote{Holds under the assumption that $\{\mybar{v}^k_n\}$ are sub-Gaussian random variables.}, and
{\setlength{\abovedisplayskip}{3pt}\setlength{\belowdisplayskip}{3pt}
\begin{equation*}
\widehat{\sigma}_{\bar{v}^k,n}^2\triangleq\frac{1}{n-1}\sum_{\ell=1}^{n-1}\left(\widehat{\mybar{v}}^k_{\ell}\right)^2,\;\; \forall n\geq 2.
\end{equation*}}
Here, $N_s\in\Nset^+$ is the length of a fixed stabilization time-interval, wherein \eqref{ourdetector} is still not sufficiently accurate, and we enforce a simple, more conservative condition for the transition phase $n\leq N_s$ (e.g., similarly to Eq.\ (49) in \cite{weiss2021blind}). The justification for \eqref{ourdetector} is omitted due to space limitation, though a similar derivation appears in \cite{weiss2021blind}, Subsection IV-C. Thus, the error propagation prevention mechanism is implemented by
\begin{equation}\label{errorpreventionrule}
\widehat{\mathbbm{1}}_{\mathcal{E}_{\tiny {\rm{OL}}_n}}=1 \Rightarrow \text{Reset:}\;\alpha_{n+1}=\alpha_0,\; \widehat{\H}^p_{n+1}=(\alpha_0/\alpha_{n})\cdot\widehat{\H}^p_{n}.
\end{equation}

To Summarize, our proposed algorithm for the implementation of the parallel mod-ADCs is executed by iteratively repeating Phases 1 and 2, where after each call to {\texttt{BlindSTModUnfold}} (Algorithm \ref{Algorithm2}) in the repeated Phase 1, we check \eqref{ourdetector} and reset the resolution if necessary \eqref{errorpreventionrule}.

\begin{figure}[t]
	\includegraphics[width=0.485\textwidth]{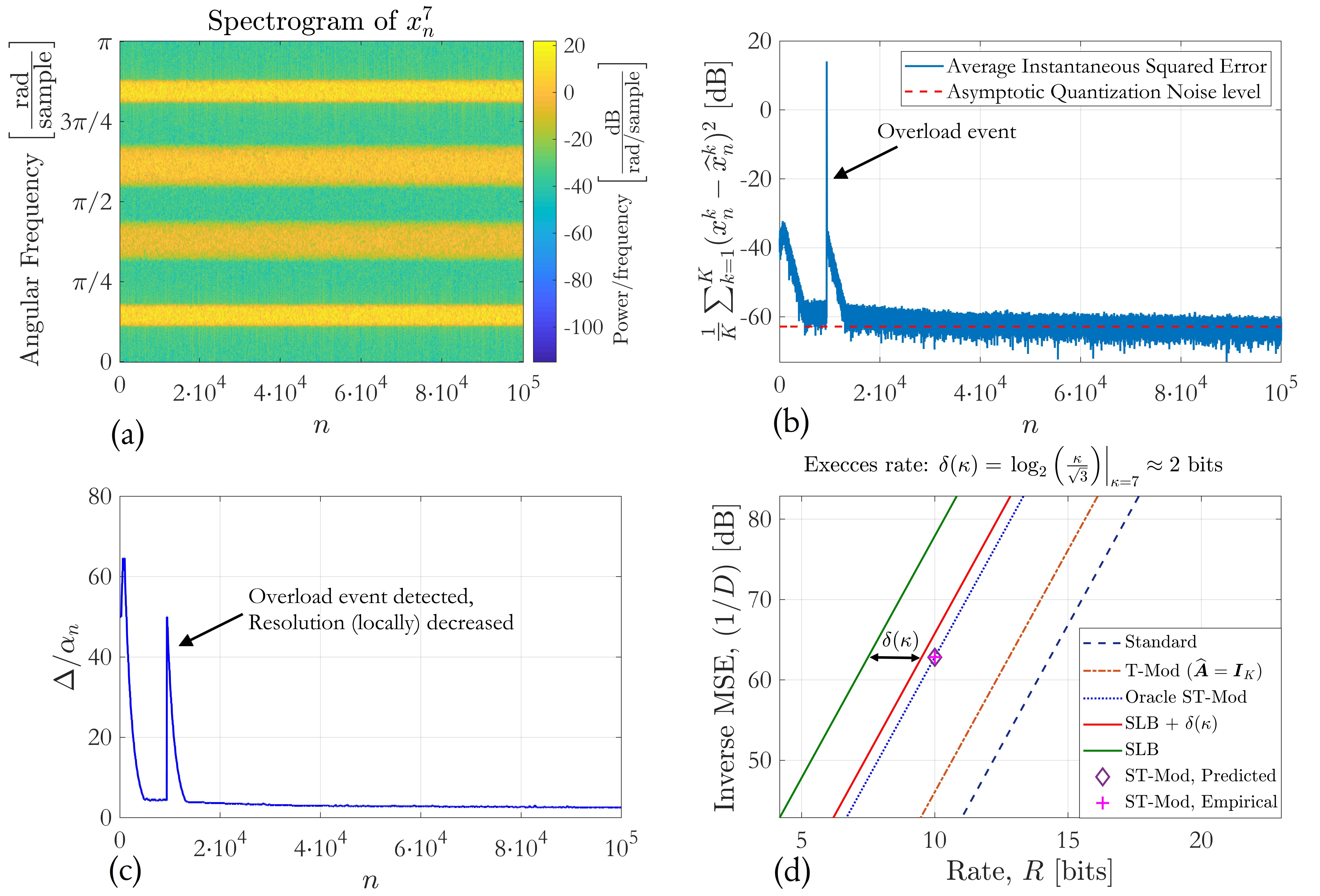}
	\centering
	\caption{Spectrogram of the input, simulation results, and rate-distortion curves.}
	\label{fig:simulationresults}
\end{figure}
\vspace{-0.25cm}
\section{Simulation Results}\label{sec:results}
\vspace{-0.2cm}
We consider $K=10$ parallel mod-ADCs, operating blindly according to the algorithm described above, whose inputs are the noisy mixtures,
\begin{equation*}
    \ux_n=\mGamma\us_n + \uxi_n\in\Rset^{K\times 1},
\end{equation*}
where $\us_n\in\Rset^{K_s\times 1}$ is a collection of $K_s=4$ zero-mean, unit-variance spectrally-flat uncorrelated Gaussian bandlimited sources, $\mGamma\in\Rset^{K\times K_s}$ represents the (spatial) channel response, and $\uxi_n$ is an additive white Gaussian noise, with $\Eset\left[\uxi_n\uxi^{\tps}_{n-\ell}\right]=\mathbbm{1}_{\ell=0}\cdot\sigma^2\I_K$. We set $R=10$, $\alpha_0=\frac{\Delta}{5K}$, $\kappa=7$, $p=30$, $L_s=2.5p$, and the entries of $\mGamma$ were drawn independently from the standard Gaussian distribution. Other system parameters, such as $\mu$ and $N_s$, were set according to the same guiding principles presented in \cite{weiss2021blind}, Subsection IV-D. We define $\mathrm{SNR}\triangleq1/\sigma^{2}$.

For an input of length $N=10^5$, Fig.\ \ref{fig:simulationresults}(a) presents the spectrogram of $x_n^7$ (as a representative example), a noisy linear mixture of the sources $\us_n$, in which the bands of the sources are clearly seen. Fig.\ \ref{fig:simulationresults}(b) presents the average instantaneous squared error in dB of $\widehat{\ux}_n$, attained by our blind mod-ADCs algorithm for $\mathrm{SNR}=30$dB. The empirical error probability in perfectly recovering $\uv_n$ is $\widehat{\Pr}(\uv_n\neq\widehat{\uv}_n)\triangleq\frac{1}{N}\sum_{\ell=1}^{N}\mathbbm{1}_{\uv_n\neq\hat{\uv}_n}=0.001\%$, which allows for accurate estimation of $\ux_n$, asymptotically matching the limiting quantization noise level (governed by the resolution parameter $\alpha_n$). Fig.\ \ref{fig:simulationresults}(c) presents the effective modulo size $\Delta/\alpha_n$, which is inversely proportional to the resolution parameter. It is seen that due to successful learning of the optimal filter $\widehat{\H}_n$ and the decoding matrix $\widehat{\A}$, the resolution parameter gradually increases, and converges to an asymptotic value near the theoretical limit for which $\kappa\cdot\sigma_{\max}=\frac{\Delta}{2}$ (see \eqref{indicatorkappastd}), where $\sigma_{\max}$ is a function of the resolution parameter. Lastly, Fig.\ \ref{fig:simulationresults}(d) presents the rate-distortion curves of standard ADCs, individual blind mod-ADCs, exploiting only temporal correlations (T-Mod), which corresponds to the special case $\widehat{\A}=\I_K$ (i.e., without IF decoding), and the predicted, as well as the empirical, rate-distortion operating point of the proposed spatiotemporal mod-ADCs with $R=10$ at the asymptotic value of $\alpha_n$ (ST-Mod). We also present the curve of the oracle decoder (Algorithm \ref{Algorithm1}) and Shannon's Lower Bound (SLB, \cite{ordentlich2018modulo}) as theoretical benchmarks. Our blind method achieves the oracle performance, and the gap from SLB is governed mainly by $\kappa$, controlling the overload probability as implied by \eqref{indicatorkappastd} (see also \cite{weiss2021blind}, Subsection IV-E), with a small gap due to IF sub-optimality. The substantial gain relative to the competing methods is evident.

\vspace{-0.35cm}
\section{Conclusion}\label{sec:conclusion}
\vspace{-0.2cm}
In the context of analog-to-digital conversion, we presented an adaptive algorithm, allowing for a stable and reliable blind operation of $K$ parallel mod-ADCs, i.e., without access to prior knowledge of the input signals. Specifically, we extended our recent work, in which only temporal correlations were exploited, to the more general case where spatiotemporal correlations are utilized for accurate estimation and subsequent decoding. We demonstrated in simulation the successful operation of the proposed solution, corroborating our derivation and the underlying theory.

\bibliographystyle{IEEEbib}
\small{\bibliography{refs}}

\end{document}